\documentstyle[12pt,doublespace,psfig]{article}

\def\ppap{
\mathrel{\hbox{\rlap{\hbox{\lower4pt\hbox{$\sim$}}}\hbox{$<$}}}}
\def\qqap{
\mathrel{\hbox{\rlap{\hbox{\lower4pt\hbox{$\sim$}}}\hbox{$>$}}}}

\newcounter{ref}%
\newcommand{\bib}{\refstepcounter{ref}{\sl\arabic{ref}}}
\newcommand{\wbib}[1]{{\ref{#1}.}~}
\def\Msol{\hbox{$\hbox{M}_\odot$}}
\def\Mjup{\hbox{$\hbox{M}_{\rm Jup}$}}

\def\v@lp@ge:#1,#2,#3:{ {\bf#1},~#2~(#3).}
\def\volumeandpage#1{\v@lp@ge:#1:}
\def\journal#1#2{{\it #1}~\volumeandpage{#2}}
\def\aaa #1{\journal{Astron. Astrophys.}{#1}}
\def\aaal #1{\journal{Astron. Astrophys. (Letters)}{#1}}

\def\aj #1{\journal{Astron. J.}{#1}}
\def\apj #1{\journal{Astrophys. J.}{#1}}
\def\apjl #1{\journal{Astroph. J. (Letters)}{#1}}
\def\apjs #1{\journal{Astroph. J. Suppl.}{#1}}

\def\aspcs #1{\journal{Asp. Conf. Ser.}{#1}}
\def\baas #1{\journal{Bulletin of the AAS.}{#1}}

\def\nature #1{\journal{Nature}{#1}}

\def\science #1{\journal{Science}{#1}}

\vspace*{-1cm}

\begin{document}

\centerline{\Large \bf Discovery of a Low-Mass Brown Dwarf} 
\vspace*{1cm}
\centerline{\Large \bf Companion of the Young Nearby Star G\,196-3}
\vspace*{1cm}
\centerline{\bf Rafael~Rebolo$^{\ast}$, Mar\'\i a~R.~Zapatero~Osorio, 
            Santiago~Madruga}
\centerline{\bf V\'\i ctor~J.~S.~B\'ejar, Santiago~Arribas and Javier~Licandro}
\vspace*{1cm}
\centerline{Instituto de Astrof\'\i{}sica de Canarias, 
             E--38200 La Laguna, Tenerife, Spain }
\vspace*{1cm}
\centerline{{\sl e-mail addresses:} \ \ rrl@ll.iac.es, mosorio@ll.iac.es, 
                                        madruga@ll.iac.es}
\centerline{\ \ \ \ vbejar@ll.iac.es, sam@ll.iac.es, jlicandr@ll.iac.es}

\vfil
\centerline{Revised version (October 1998)}

\vfil
$^{\ast}$ {To whom correspondence should be addressed; 
           {\sl e-mail:} rrl@ll.iac.es}

\vfil\eject
\pagestyle{myheadings}
\markboth
{{\scriptsize R. Rebolo et al. \ \ \ Discovery of a Low-Mass Brown Dwarf 
Companion of a Young Nearby Star}}
{{\scriptsize R. Rebolo et al. \ \ \ Discovery of a Low-Mass Brown Dwarf 
Companion of a Young Nearby Star}}

\noindent
{\bf 
A substellar-mass object in orbit at $\sim$ 300 astronomical units (AU) 
from the young low-mass star G\,196-3 was detected by direct imaging. 
Optical and infrared photometry and low- and intermediate-resolution 
spectroscopy of the faint companion, hereafter referred to as G\,196-3B, 
confirms its cool atmosphere and allows its mass to be estimated at
25$^{+15}_{-10}$ Jupiter masses (\Mjup). The separation between both objects 
and their mass ratio suggest the fragmentation of a collapsing cloud as the 
most likely origin for G\,196-3B, but alternatively it could have originated 
from a proto-planetary disc which has been dissipated. Whatever the formation 
process was, the young age of the primary star ($\sim$100\,Myr) demonstrates 
that substellar companions can form in short time scales.}

\newpage

Direct imaging searches for brown dwarfs and giant planets around
stars explore a range of physical separations complementary to that of
radial velocity measurements and provide key information on how
substellar-mass companions are formed. Any companion uncovered by an
imaging technique can be further investigated by spectroscopy which
allows information to be obtained on its atmospheric
conditions and evolutionary status. So far, only one unambiguous brown
dwarf companion to a star has been imaged (\bib{\label{nakajima95}}) 
and subsequently investigated in detail (\bib{\label{oppenheimer95}}, 
\bib{\label{oppenheimer98}}, \bib{\label{schultz98}}). 
Young, nearby cool dwarf stars are ideal search targets for
substellar-mass companions (brown dwarfs and giant planets) using
direct imaging techniques because: {\sl (i)} young substellar objects 
are considerably more luminous when undergoing the initial phases of 
gravitational contraction (\bib{\label{dantona94}}, \bib{\label{baraffe95}}, 
\bib{\label{burrows97}}) than at later stages; {\sl (ii)} stars in the 
solar neighbourhood (i.e. within 50\,pc of the Sun) allow the detection 
of faint companions at physical separations of several tens of AU; 
and {\sl (iii)} cool stars are among 
the least luminous stars which favor full optimization of the dynamical 
range of current detectors to achieve detection of companions as faint 
as possible by means of narrow-band imaging techniques at red wavelengths.

Using X-ray emission as an indicator of youth (\bib{\label{note1}}, 
\bib{\label{hempelman95}}, \bib{\label{randich96}}) we have selected a 
number of late-type stars (K and M spectral classes) in the solar 
neighbourhood for which we have obtained deep images (down to a limit of 
about 19\,mag in $I$) in narrow-band filters (\bib{\label{martin96}}) 
centered at 740 and 914\,nm (with a
bandwidth of 10\,nm). The survey is being conducted at the 0.8-m 
telescope of the Instituto de Astrof\'\i sica de Canarias (IAC80)
at the Teide Observatory (OT) on Tenerife using the Thomson
1024$\times$1024 charge-coupled device (CCD). One pixel of this detector
projects 0.432\,arc\,sec on the sky.  The two narrow filters allow
effective discrimination of faint red objects at separations
larger than 3--4 times the full width at half maximum of the point
source response which, for the first 52 targets of the program, was
on average close to 1.5\,arc\,sec. Here we report on the
discovery of a very red companion to the high proper motion M-dwarf
star G\,196-3. The observations were performed on 25 January 1998.  
A comparison of the images taken at different wavelengths
showed that a faint red companion was present at 16.2\,arc\,sec
southwest of the star (position angle\,=\,210\,$^{\circ}$, see Fig.~1). 
We have named this companion as G\,196-3B. 
Further optical $R$ and $I$ broad-band photometry (Table~1) was obtained 
at the 1-m Optical Ground Station (OGS) telescope on 19 March 1998, while 
infrared $J$ and $K$ data (Table~1) were collected at the 1.5-m Carlos 
S\'anchez Telescope (TCS) on 24 March 1998, both telescopes located at the 
OT.  The colors of G\,196-3B clearly reveal that it is one 
of the reddest objects known.

Inspection of the second Palomar Observatory Sky Survey (POSS-2) red 
plates (obtained from the STScI Digitized Sky Survey) provides a 2-$\sigma$ 
detection of G\,196-3B at the position expected for a proper
motion (\bib{\label{giclas71}}) common with that of G\,196-3. 
Images in the $I$-band taken
with ALFOSC at the 2.5-m Nordic Optical Telescope (NOT) at the Roque de
los Muchachos Observatory (ORM) on 16 February 1998 (with a pixel size
of 0.187\,arc\,sec) and with HIRAC on 3 June 1998 (pixel size
0.109\,arc\,sec) confirm that the faint object has a proper motion 
($\mu_{\alpha{\rm cos}\delta}$\,=\,--0.5$\pm$0.1\,arc\,sec\,yr$^{-1}$,
$\mu_\delta$\,=\,--0.3$\pm$0.1\,arc\,sec\,yr$^{-1}$)
consistent with that of the M star within 2-$\sigma$ error bars 
(\bib{\label{note2}}). 

A low-resolution optical spectrum of G\,196-3B (Fig.~2, top panel) was 
obtained at the NOT on 16 February 1998, using the low-resolution ALFOSC
spectrograph, grating no.~4 (dispersion 3.2\,\AA\ per pixel, effective
resolution 16\,\AA) and a LORAL 2048$\times$2048 detector. It shows 
distinctive features which are characteristic of temperatures cooler 
than 2000\,K. G\,196-3B's spectrum is dominated by two pronounced 
absorptions centered 
on 769 and 868\,nm. Both features have recently been identified
as due to a considerable broadening of the 766.4, 769.8\,nm K\,{\sc i}
resonance doublet (\bib{\label{martin97}}, \bib{\label{tinney98}}, 
\bib{\label{pavlenko98}}) and due to molecular absorptions of CrH 
(\bib{\label{kirkpatrick98}}) at 861.1 and 869.6\,nm and FeH 
({\sl\ref{kirkpatrick98}}) at 869.2\,nm, respectively. 
TiO and VO molecular absorptions, which are
very intense in M-type dwarfs, appear weak or non-existent in our
spectrum possibly indicating the photospheric depletion of Ti and V
atoms  to form dust grains. All this, together with the comparison of
our data with the spectra of similar resolution obtained for
Kelu\,1 (\bib{\label{ruiz97}}) and DENIS
objects (\bib{\label{delfosse97}}), suggests that G\,196-3B deserves
to be considered as an appropriate candidate for the so-called L-type
class ({\sl\ref{martin97}}). There is no evidence for H$_{\alpha}$ in
emission, and the atomic lines of Na\,{\sc i}, Rb\,{\sc i} and Cs\,{\sc
i}, which are observed in other L-type objects, show up 
weakly in G\,196-3B. Their weakness can be interpreted as an indication
of low surface gravity consistent with the expectation of a very young
object undergoing gravitational contraction. From the similarity of our 
spectrum to that of Kelu\,1 we infer an effective temperature 
of 1800$\pm$200\,K for G\,196-3B. The probability 
of finding a free-floating L-type field object in our survey 
(0.74\,deg$^2$) is rather small ($\le$4\%) given the statistics 
of detections of similar objects achieved by the 
large-scale infrared surveys 
DENIS ({\sl\ref{delfosse97}}) and 2MASS (\bib{\label{beichman98}}). 
This provides additional support for G\,196-3B being a genuine 
companion to the M primary star.

A low-resolution optical spectrum of G\,196-3 covering a wavelength 
range of 449--900\,nm was obtained at the NOT  with the same instrumental 
setup and on the same date as that of the companion. Using spectral
indicators (\bib{\label{prosser91}}, \bib{\label{martin94}}) based on
the strength of several TiO bands present in this wavelength interval,
we classify our star as an  M2.5 dwarf with an uncertainty of half a
subclass. Broad-band $VRIJHK$ photometry (Table~1) obtained with
the IAC80 telescope on 27 February 1998 and with the TCS telescope on 
17--20 June 1998, agrees with this spectral classification. The optical 
spectrum is similar to that of M2--3 dwarfs in young stellar clusters of 
solar metallicity. The observed optical and infrared colors present no 
strange anomaly that might be attributed to an unresolved less massive 
companion to G\,196-3, and no indication of changes in the radial velocity 
is found beyond the uncertainties of our measurements ($\pm 6$\,km\,s$^{-1}$) 
determined using high-resolution spectra taken at the INT over a time 
interval of several hours to days. This makes very unlikely that the star 
is actually a close contact binary. The spectral type combined with the 
observed fluxes indicates the star is at a minimum distance of 15.4\,pc 
from Earth; this distance would correspond to the star's being at the 
hydrogen-burning stage on the main sequence. As we shall discuss below, 
there are reasons to believe the star is young and in a pre-main sequence 
phase; it is therefore at a more luminous evolutionary stage and 
consequently at a greater distance.

H$_{\alpha}$ and H$_{\beta}$ are seen in emission with equivalent widths 
(EWs) of 0.40$\pm$0.03\,nm and 0.33$\pm$0.03\,nm, respectively, indicating a
significant chromospheric activity in G\,196-3; this is also 
supported by the strong emission of Ca\,{\sc ii} H and K (EWs of 1.7 and 
1.0$\pm$0.1\,nm, respectively) and other Balmer lines 
(\bib{\label{polomski96}}) (see Fig.~3), as well as by the star 
being an extreme ultraviolet source (\bib{\label{lampton97}}). 
This when taken together with the high X-ray flux 
(48$\times$10$^{-13}$\,erg\,cm$^{-2}$\,s$^{-1}$) measured
by the ROSAT All Sky Survey (RASS), which at the minimum distance of
the star results in a X-ray luminosity of $L_{\rm X} = 
1.36\times10^{29}$\,erg\,s$^{-1}$, supports the hypothesis that
we are dealing with a young star. G\,196-3 shows the same
chromospheric and coronal properties of stars of similar temperatures
in the young open cluster $\alpha$\,Persei (60\,Myr) 
(\bib{\label{basri98}}) and slightly more activity than the average of 
its counterparts in the Pleiades (120\,Myr) (\bib{\label{martin98}}). 
The coronal and chromospheric properties, when examined in the 
light of the available data on the activity of stars of similar spectral 
type in these clusters ({\sl\ref{hempelman95}, \ref{randich96}}), suggest 
an age for G\,196-3 of $\sim$100\,Myr, i.e. intermediate between that of 
the $\alpha$~Persei and Pleiades clusters.

An upper limit to the age can be imposed from the comparison to the Hyades 
cluster (600\,Myr), where the average of chromospheric and coronal emission 
of M2--M3 stars is considerably lower than in G\,196-3. This star appears 
to be significantly younger than the Hyades and hence, we will adopt 
300\,Myr, an age intermediate between that of the Pleiades and Hyades, 
as a reasonable age upper limit. The lower age limit can be derived from 
observations of Li\,{\sc i} at 670.8\,nm. Lithium is a fragile element which 
burns efficiently in the interiors of fully convective stars over short time 
scales (a few tens of Myr).
Convection drains material from the stellar atmosphere into the
innermost  layers where the  temperature is high enough for Li burning
to take  place. There are several
models ({\sl\ref{dantona94}, \ref{baraffe95}}) in the literature
that predict the Li depletion rate as function of mass for low-mass
stars and give consistent results.  A search was made for the Li\,{\sc
i} line in G\,196-3, and an optical spectrum was obtained on 13
February 1998 using the Intermediate Dispersion Spectrograph with the
235-mm camera and a 1024$\times$1024 CCD on the INT at the ORM.  The
H1800V grating gave a nominal dispersion of 0.053\,nm per pixel and an
effective resolution of 0.1\,nm. The spectral range covered was
640--696.5\,nm. We imposed an upper limit on the EW of
0.005\,nm which gives a Li depletion factor larger than 1000 with
respect to its original abundance. This constrains the age of the star
to be older than 20\,Myr ({\bib{\label{chabrier96}}). 
All these considerations provide a most likely
age for G\,196-3 that locates our star in the pre-main sequence
evolutionary phase, and thus at a more luminous stage than expected for
its main-sequence lifetime.  According to the age range derived, the
most probable distance to the system is 21$\pm$6\,pc, the minimum value
corresponding to the case of the primary star already on the
main sequence and the maximum distance taking into account the youngest
possible age.

Assuming this distance interval, we can estimate the luminosity of the
companion, G\,196-3B, from the measured $I$ and $K$ magnitudes and the
$K$ bolometric correction (\bib{\label{bessell98}}) as a function of
the color $(I-K)$. The values obtained are 
log\,$L/L_{\odot}$\,=\,--4.1 conservatively assuming the oldest age 
(main-sequence) and log\,$L/L_{\odot}$\,=\,--3.6 for the youngest
one. The comparison of our optical/infrared magnitudes 
with the recent evolutionary tracks (\bib{\label{burrows98}}, 
\bib{\label{baraffe98}}), which include dust condensation, allows us 
to conclude that the mass of G\,196-3B is 25$^{+15}_{-10}$\,\Mjup, where 
the upper and lower values result from the age limits discussed above.

An independent confirmation of the substellar nature of this faint
companion was achieved with the detection of the Li\,{\sc i}
resonance doublet at 670.8\,nm. We obtained an intermediate-resolution
optical spectrum at the 4.2-m William Herschel Telescope (WHT) at the
ORM on 6 April 1998 (Fig.~2, bottom panel) using the ISIS double-arm
spectrograph, grating of 316 grooves mm$^{-1}$ and a 1024$\times$1024
Tektronix CCD, an instrumental setup which provided a nominal
dispersion of 0.15\,nm per pixel (an effective resolution of 0.3\,nm).
The EW of the doublet is 0.5$\pm$0.1\,nm which, using model
atmospheres ({\sl\ref{pavlenko98}}), gives an atmospheric
abundance consistent with no depletion  at all of lithium. The
presence of Li combined with the low atmospheric temperature 
rules out the possibility that our object is a star. Any brown dwarf
with  a mass below 65\,\Mjup \ should preserve its initial Li content
for its entire lifetime, and an object with such a small mass as that
of  G\,196-3B should necessarily show a high Li content. 
While in more massive substellar
objects the presence of Li would help to determine its
evolutionary stage  more precisely through the time dependence of Li
burning, for our object this detection provides a necessary check of
consistency.

A more precise evolutionary status of G\,196-3B can be determined 
with an accurate parallax measurement and/or the detection of deuterium
in the atmosphere. Deuterium, a more fragile nuclear species than
Li, is burnt ({\sl\ref{dantona94}}, \bib{\label{saumon96}})
in substellar-mass objects with masses above 12--15\,\Mjup. The consumption
of deuterium provides a significant part of the total luminosity of
brown dwarfs during the early phases that takes from 1 to 100\,Myr
depending on its mass and age. G\,196-3B has probably burnt its
deuterium, but it could have preserved a detectable amount  if its mass
were indeed close to 15\,\Mjup. Detection of deuterium features
(deuterated molecules, hyperfine-transition isotopic splitting, etc.)
could contribute to better determine the  mass and evolutionary status 
of G\,196-3B.
}
 
The distance of the system implies a physical separation between the
two components of more than 250 AU, being 350\,AU at 21\,pc. 
It could be even larger if the system were younger and 
therefore more distant from the Sun. This large distance and the high mass 
ratio of 16:1 between the two components favors the fragmentation of a
collapsing cloud as the most plausible explanation for the formation of
the system (\bib{\label{artymowicz98}}, \bib{\label{bodenheimer98}}).  It
cannot be excluded, however, that the accretion of matter in a
proto-planetary disc may produce an object more massive than 15\,\Mjup \ 
at such large distances. Accretion discs extending up to several
hundred astronomical units are known to exist around several 
stars (\bib{\label{bruhweiler97}}). Similar surveys to that conducted 
here will
provide a statistically significant number of substellar-mass
companions to test the proposed formation mechanisms and may well
promote the development of new ideas as occurred with the recent
findings of giant planets with highly eccentric orbits around
solar-type stars (\bib{\label{mayor95}}, \bib{\label{marcy98}}).  Of
the 52 stars we have examined so far, only G\,196-3 has
shown a substellar companion at distances larger
than 60\,AU, the minimum physical distance that we can explore given
the characteristics of our survey. We may infer from this that the
percentage of stars with substellar companions of roughly this mass at
this or larger distances may be of the order of 2\%. This is
similar to the number of stars that show such companions at
distances smaller than 5\,AU according to searches based on
radial-velocity
measurements ({\sl\ref{mayor95}, \ref{marcy98}}).

\newpage

\noindent
\hrule

\noindent
References

\footnotesize
\noindent
\begin{itemize}
\item[\wbib{nakajima95}] T. Nakajima et al., \nature{378,463,1995}

\item[\wbib{oppenheimer95}] B. R. Oppenheimer, S. R. Kulkarni, K. Matthews,  
      T. Nakajima, \science{270,1478,1995}

\item[\wbib{oppenheimer98}] B. R. Oppenheimer, S. R. Kulkarni, K. Matthews, 
      M. H. van Kerkwijk, {\sl Astrophys. J.}, submitted (1998).

\item[\wbib{schultz98}] A. B. Schultz et al., \apj{492,L181,1998}

\item[\wbib{dantona94}] F. D'Antona \& I. Mazzitelli, \apjs{90,467,1994}

\item[\wbib{baraffe95}] I. Baraffe, G. Chabrier, F. Allard, P. Hauschildt,
      \apjl{446,L35,1995}

\item[\wbib{burrows97}] A. Burrows et al., \apj{491,856,1997}

\item[\wbib{note1}] X-ray emission of low-mass stars is associated to the 
magnetic activity resulting from the interplay between stellar rotation 
and surface convective motions. The stellar rotation is a decreasing 
function of age in these stars. Young low-mass stars typically show rapid 
rotation which favors the generation of enhanced magnetic fields and the 
heating of the outer-most stellar layers producing the X-ray emission.

\item[\wbib{hempelman95}] A. Hempelmann, J. H. M. M. Schmitt, M. Schultz, 
      G. R\"udiger,  K. Stepi\`en, \aaa{294,515,1995}

\item[\wbib{randich96}] S. Randich, J. H. M. M. Schmitt, C. F. Prosser, 
      J. R. Stauffer, \aaa{305,785,1996}

\item[\wbib{martin96}] E. L. Mart\'\i n, R. Rebolo, 
      M. R. Zapatero Osorio, \apj{469,706,1996}

\item[\wbib{giclas71}] $\mu_{\alpha{\rm cos\delta}}$\,=\,--0.2
      $\pm$0.1\,arc\,sec\,yr$^{-1}$, $\mu_\delta$\,=\,--0.2
      $\pm$0.1\,arc\,sec\,yr$^{-1}$; 
      H. L. Giclas, R. Jr. Burnham, N. G. Thomas, 
      {\sl The G Numbered Stars} (Lowell Observatory at Flagstaff, Arizona, 
      1971).

\item[\wbib{note2}] Additionally, high-resolution spectra of both the 
primary star, G\,196-3, and the faint companion, G\,196-3B, obtained at 
the Keck 10-m telescope (Mauna Kea Observatory, Hawaii) by E. L. Mart\'\i n 
and G. Basri show that the radial velocity of the two components does 
coincide within 2\,km\,s$^{-1}$. 

\item[\wbib{martin97}] E. L. Mart\'\i n, G. Basri, X. Delfosse, T. 
      Forveille, \aaal{327,L29,1997}

\item[\wbib{tinney98}] C. G. Tinney, X. Delfosse, T. Forveille, F. Allard, 
      {\sl Astron. Astrophys.}, submitted (1998).

\item[\wbib{pavlenko98}] Ya. Pavlenko, M. R. Zapatero Osorio, R. Rebolo, 
      paper presented at the {\sl Very Low-Mass Stars and Brown Dwarfs in 
      Stellar Clusters and Associations} Euroconference, La Palma, Spain, 
      11--15 May 1998.

\item[\wbib{kirkpatrick98}] D. Kirkpatrick et al., {\sl Astrophys. J.}, 
      submitted (1998).

\item[\wbib{ruiz97}] M. R. Ruiz, S. K. Leggett, F. Allard, 
      \apjl{491,L107,1997}

\item[\wbib{delfosse97}] X. Delfosse et al., \aaal{327,L25,1997}

\item[\wbib{beichman98}] C. Beichman et al., paper presented at the 
      {\sl Very Low-Mass Stars and Brown Dwarfs in 
      Stellar Clusters and Associations} Euroconference, La Palma, Spain, 
      11--15 May 1998.

\item[\wbib{prosser91}] C. F. Prosser, J. R. Stauffer, R. P. Kraft, 
      \aj{101,1361,1991}

\item[\wbib{martin94}] E. L. Mart\'\i n, R. Rebolo, A. Magazz\'u, 
      \apj{436,262,1994}

\item[\wbib{polomski96}] E. Polomski, S. Vennes, J. R. Thorstensen, M. 
      Mathioudakis, E. E. Falco, \apj{486,179,1997}

\item[\wbib{lampton97}] M. Lampton, R. Lieu, J. H. M. M. Schmitt, S. Bowyer, 
      W. Voges, J. Lewis, X. Wu, \apjs{108,545,1997}

\item[\wbib{basri98}] G. Basri \& E. L. Mart\'\i n, {\sl Astrophys. J.}, 
      in press (1998).

\item[\wbib{martin98}] E. L. Mart\'\i n et al., \apjl{499,L61,1998}

\item[\wbib{chabrier96}] G. Chabrier, I. Baraffe, B. Plez, \apjl{459,L91,1996}

\item[\wbib{bessell98}] M. S. Bessell, F. Castelli, B. Plez, 
      \aaa{333,231,1998}

\item[\wbib{burrows98}] A. Burrows et al., \aspcs{134,354,1998}

\item[\wbib{baraffe98}] I. Baraffe, G. Chabrier, F. Allard, P. H. 
      Hauschildt, {\sl Astron. Astrophys.}, in press (1998).

\item[\wbib{saumon96}] D. Saumon, W. B. Hubbard, A. Burrows, T. Guillot, 
      J. I. Lunine, G. Chabrier, \apj{460,993,1996}

\item[\wbib{artymowicz98}] P. Artymowicz, \aspcs{134,152,1998}

\item[\wbib{bodenheimer98}] P. Bodenheimer, \aspcs{134,115,1998}

\item[\wbib{bruhweiler97}] F. Bruhweiler et al., \baas{191,47.03,1997}

\item[\wbib{mayor95}] M. Mayor \& D. Queloz, \nature{378,355,1995}

\item[\wbib{marcy98}] G. Marcy, \nature{391,127,1998}

\item[ ]ACKNOWLEDGEMENTS. We thank N. Gonz\'alez P\'erez for assistance with
the infrared observations at the TCS and J. Casares and I.
Mart\'\i nez-Pais for their assistance at the INT. We are also indebted to
I. Baraffe and the Lyon group for providing us with new evolutionary models 
for very low masses prior to publication; and to E. Polomski for sending 
the published spectrum of G\,196-3 in digitalized format. We are also 
indebted to E. L. Mart\'\i n and G. Basri for sharing data prior to 
publication. This paper is based on observations made
with the IAC80, the European Space Agency OGS and the TCS, operated by 
the Instituto de Astrofisica de Canarias (IAC) at the OT; and the NOT, 
the INT and the WHT  operated on the island of La Palma by the Isaac 
Newton Group at the ORM of the IAC. Partial financial support was 
provided by the Spanish DGES project no. PB95-1132-C02-01.

\end{itemize}

\baselineskip=.99\baselineskip \vfil\eject\noindent 

\normalsize

\begin{table}
\caption[]{Data for the G\,196-3 system}
\begin{center}
\begin{tabular}{c|c|c}
\hline
                  &                &                \\
                  & {\bf G\,196-3} & {\bf G\,196-3B} \\
                  &                &                \\
\hline
                  &                &                \\
$V_{\rm c}$       & 11.75$\pm$0.01 &   \\
$R_{\rm c}$       & 10.67$\pm$0.01 & 20.78$\pm$0.10 \\
$I_{\rm c}$       &  9.41$\pm$0.01 & 18.28$\pm$0.05 \\
$J_{\rm UKIRT}$   &  8.12$\pm$0.02 & 14.73$\pm$0.05 \\
$H_{\rm UKIRT}$   &  7.47$\pm$0.02 &  \\
$K_{\rm UKIRT}$   &  7.29$\pm$0.02 & 12.49$\pm$0.10 \\
SpT               & M2.5$\pm$0.5   & L-class \\
$T_{\rm eff}$ (K) & 3400$\pm$100   & 1800$\pm$200 \\
log\,$L/L_{\odot}$ & --1.5$\pm$0.2 & --3.8$^{+0.2}_{-0.3}$\\
Mass              & 0.40$\pm$0.05\,\Msol& 25$^{+15}_{-10}$\,\Mjup\\
                  &                &                \\
\hline
\end{tabular}
\end{center}
\end{table}

\vspace*{10cm}

\baselineskip=.99\baselineskip
\vfil\eject\noindent
\noindent
{\bf Figure captions.}\\

{\bf Figure 1. }
$I$-band image taken at NOT ($36\,\times\,36$\,arc\,sec$^2$) 
showing the substellar companion, G\,196-3B, discovered at 16.2\,arc\,sec SW 
(position angle\,=\,210$^{\circ}$) of the young nearby red star G\,196-3.  

{\bf Figure 2.}
{\sl (Top panel)} Optical low-resolution spectrum of the substellar companion  
G\,196-3B obtained at the NOT telescope. The spectrum (normalized to unity at 
813\,nm) shows features which are distinctive of very cool temperatures 
($T_{\rm eff} \le 2000$\,K). {\sl (Bottom panel)} Intermediate-resolution 
spectrum obtained at the WHT telescope showing the lithium detection at 
670.8\,nm in G\,196-3B.

{\bf Figure 3.}
Low-resolution spectra of the M2.5 type star G\,196-3 ({\sl\ref{polomski96}}) 
showing Ca\,{\sc ii} H and K and some Balmer lines in emission in comparison 
to other young stars of similar spectral type (AP\,246 is M3 and AP\,237 is 
M4) in the $\alpha$~Persei cluster. Spectra have been normalized to unity at 
450\,nm and a constant offset of 1 in the flux axis has been added for clarity.

\newpage

%\begin{figure}
%\centerline{\psfig{figure=f9844181.ps}}
%\end{figure}

\newpage

\begin{figure}
\centerline{\psfig{figure=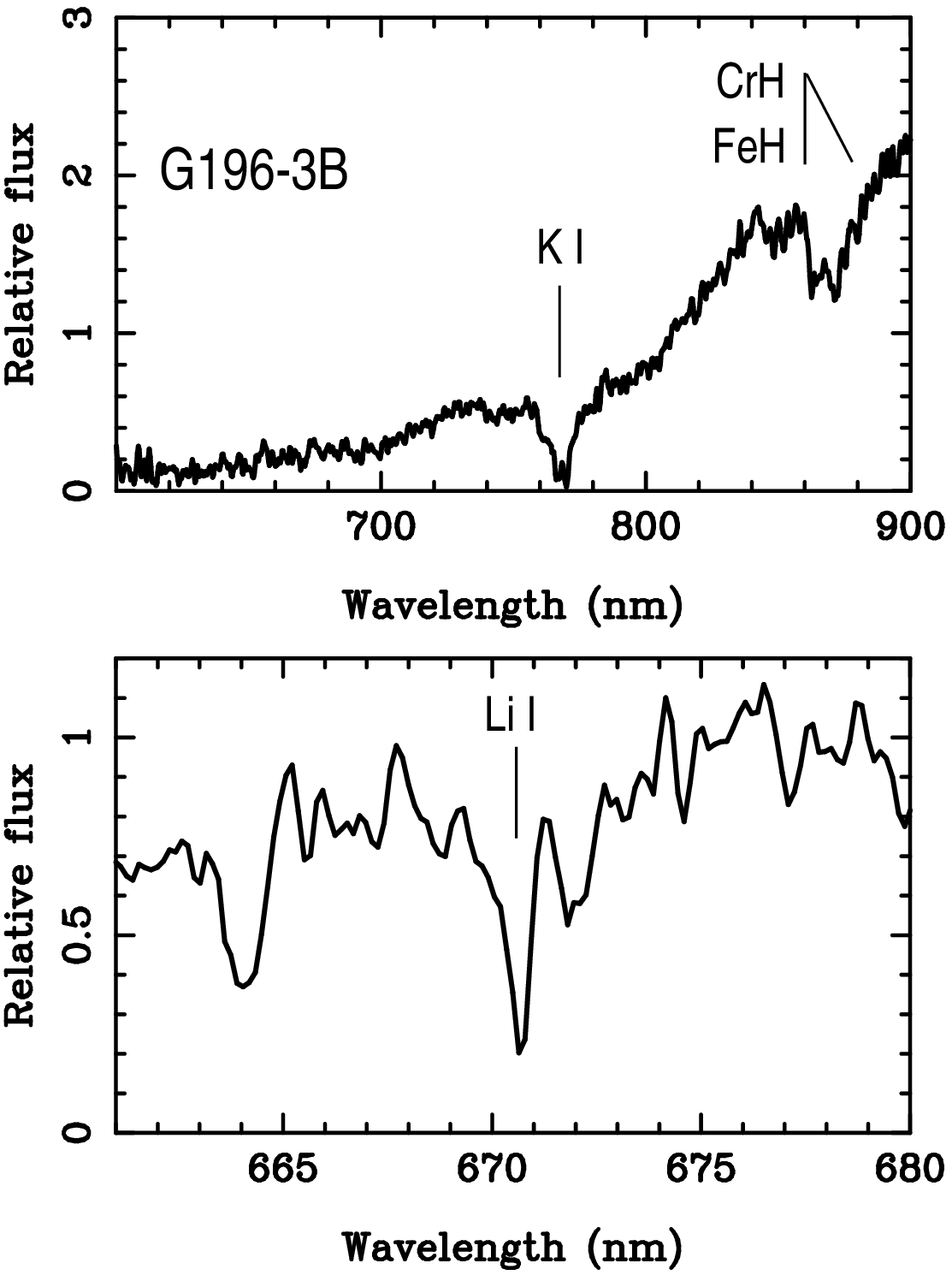}}
\end{figure}

\newpage

\begin{figure}
\centerline{\psfig{figure=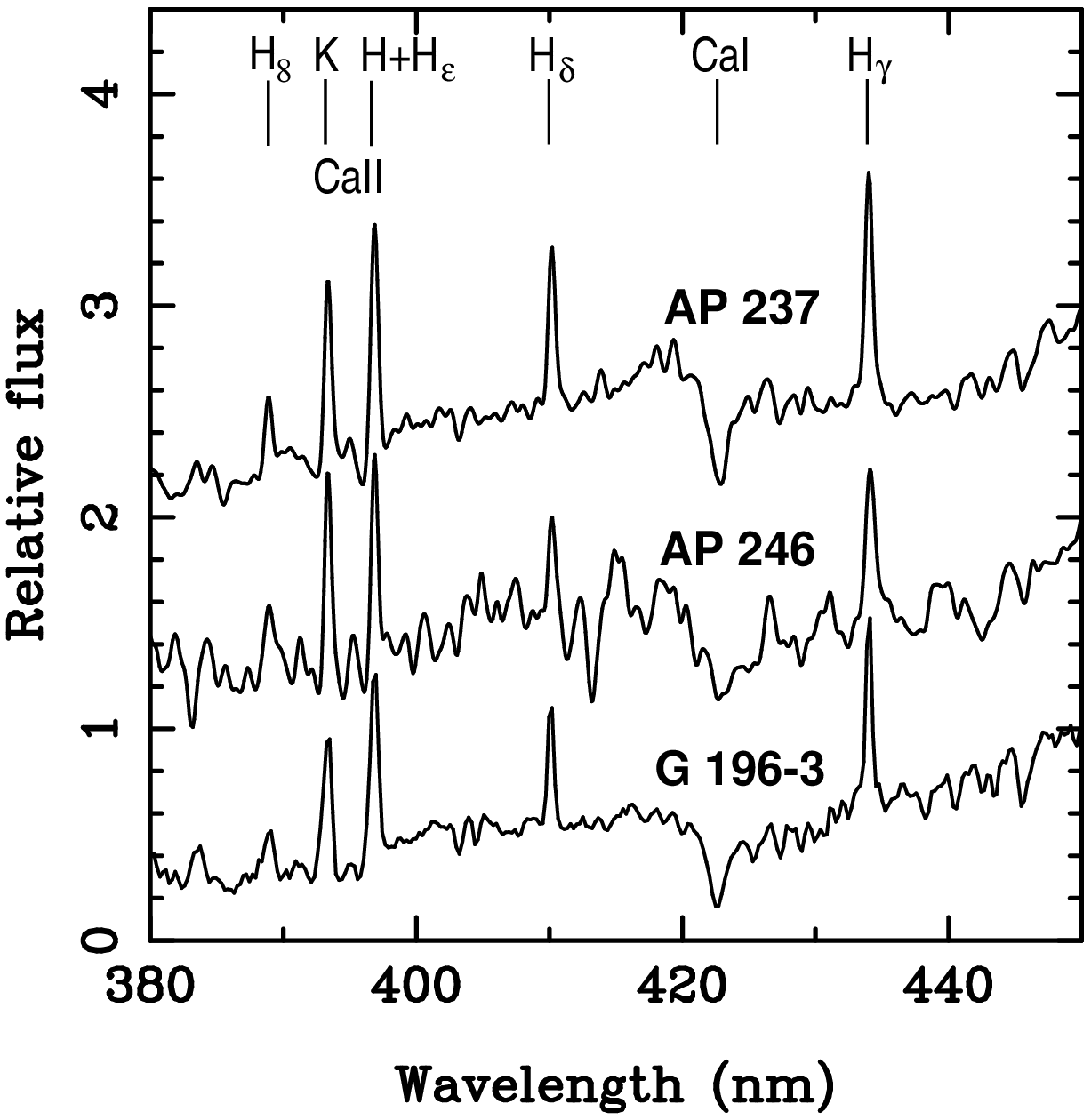}}
\end{figure}

\end{document}